\documentclass{Rinton-P9x6}

\def\rightnote{Talk given at CICHEP '2001, Cairo, Egypt, January 2001.}

\def\beq{\begin{equation}}
\def\eeq{\end{equation}}
\def\barr{\begin{eqnarray}}
\def\earr{\end{eqnarray}}
\def\lsim{\raise0.3ex\hbox{$\;<$\kern-0.75em\raise-1.1ex\hbox{$\sim\;$}}}
\def\gsim{\raise0.3ex\hbox{$\;>$\kern-0.75em\raise-1.1ex\hbox{$\sim\;$}}}
\def\lg{\raise0.3ex\hbox{$\;>$\kern-0.75em\raise-1.1ex\hbox{$<\;$}}}
\def\equiapp{\raise0.3ex\hbox{$\;\sim$\kern-0.75em\raise-1.1ex\hbox{$=\;$}}}

\def\dmsq{\Delta m^2}
\def\ue3{U_{e3}}
\def\ue3sq{|U_{e3}|^2}

\begin{document}

\title{earth matter effects on the supernova neutrino spectra}

\author{ Amol S. Dighe}

\address{Div. TH, CERN, CH-1211 Geneva 23, Switzerland\footnote{
Current address: 
Max-Planck-Institut f\"ur Physik, F\"ohringer Ring 6,
DE-80805 M\"unchen, Germany.}. \\
E-mail:  amol.dighe@cern.ch}

\maketitle

\abstracts{
We explore the earth matter effects on the energy spectra of 
neutrinos from a supernova. 
We show that the observations of the energy spectra 
of $\nu_e$ and $\bar{\nu}_e$ from a  galactic supernova 
may enable us to identify the solar neutrino solution, 
to determine the sign of $\Delta m^2_{32}$,
and to probe the mixing matrix element 
$|U_{e3}|^2$ to values as low as $10^{-3}$.
We point out scenarios in which the matter effects can  
even be established through the observation of 
the spectrum at a single detector.
}

\section{Introduction}

In the solutions of the solar and atmospheric
neutrino anomalies through the oscillations
between the three active neutrino species, 
three types of ambiguities remain to be resolved:
(i) the  solution of the solar neutrino problem,
(ii) the mass hierarchy --
normal ($m_3 > m_1, m_2$) or inverted ($m_3 < m_1, m_2)$, and
(iii) the  value of $|U_{e3}|^2$.
The energy spectra of neutrinos from a Type II supernova
contain signatures that can resolve some of these ambiguities
irrespective of the model of supernova dynamics \cite{ds}.
Here we concentrate on the effects due to the passage of 
these neutrinos through the earth which may give spectacular
signatures of some of the neutrino mass and mixing schemes.

The neutrinos from the supernova arrive at the earth as an 
effectively incoherent mixture of the mass eigenstates in
vacuum. Depending on the relative position of the supernova
and the neutrino detector at the time of arrival, the neutrinos 
have to travel different distances through the earth. 
The mass eigenstates in
vacuum then get intermixed and the observed energy spectra 
of neutrinos are affected. Since the energy range of the
supernova neutrinos overlaps with the energy range for the MSW
resonance with $\dmsq_\odot$, the matter effects 
can become large.
In that case, we have distinctive signatures of some of 
the neutrino mixing schemes.
The comparison of the neutrino spectra observed at two 
detectors can be used to measure the earth matter effects,
but in certain situations even the spectrum observed
at a single detector can perform the task.

\section{Neutrino Conversions in the star and the earth}

The neutrino transitions between different matter eigenstates
inside the supernova take place mainly in the resonance
regions $H$ and $L$, which are characterized by 
$(\Delta m^2_{atm},4|U_{e3}|^2)$ and 
$(\Delta m^2_\odot, \sin^2 2\theta_\odot)$ respectively.
Due to the $\Delta m^2$-hierarchy ($\Delta m^2_{atm} \gg
\Delta m^2_\odot$), the dynamics in 
each of the two resonance layers
can be considered independently as a $2\nu$ transition
\cite{factorize}.
The final neutrino fluxes can then 
be written in terms of {\it survival probabilities}
$p$ and $\bar{p}$ of $\nu_e$ and $\bar{\nu}_e$ respectively
(apart from the geometrical factor of $1/R^2$):
\beq
{\small
\left[ \begin{array}{c}
F_e \\ F_{\bar{e}} \\ 4 F_x
\end{array} \right] =
\left[ \begin{array}{ccc}
p & 0 & 1-p \\ 0 & \bar{p} & 1 - \bar{p} \\
1-p & 1 - \bar{p} & 2 + p + \bar{p}
\end{array} \right]
\left[ \begin{array}{c}
F_e^0 \\ F_{\bar{e}}^0 \\  F_x^0
\end{array} \right],}
\label{tr-matrix}
\eeq
where $F_e (F_e^0), F_{\bar{e}} (F_{\bar{e}}^0)$ and
$F_x (F_x^0)$ are the final (initial) fluxes of
$\nu_e, \bar{\nu}_e$ and $\nu_x$ (each of the 
non-electron neutrino or antineutrino species) respectively.

Let $P_H (\bar{P}_H)$ and $P_L (\bar{P}_L)$ be the probabilities that
the neutrinos (antineutrinos) jump to another matter eigenstate in the
resonance layers $H$ and $L$ respectively.
Adiabaticity in the resonance layer $H$ divides the possible range
of $\ue3sq$ into three regions \cite{ds,taup}: 
(I) completely adiabatic transitions, $P_H \lsim 0.1, 
\ue3sq \gsim 10^{-3}$, 
(II) partially adiabatic transitions, $0.1 \lsim P_H \lsim 0.9,
10^{-5} \lsim \ue3sq \lsim 10^{-3}$,
(III) completely adiabatic transitions, $P_H \gsim 0.9,
\ue3sq \lsim 10^{-5}$. (For inverted hierarchy, $P_H$ needs to
be replaced by $\bar{P}_H$. The ranges of $\ue3sq$ do not
change much, however.)

The values of $p$ and $\bar{p}$ are determined by these {\it flip
probabilities} and the mixing matrix elements $|U_{ei}|^2$:
\beq
p = \sum a_i |U_{ei}|^2 \quad , \quad
 \bar{p} = \sum \bar{a} |U_{ei}|^2~,
\eeq
where $a_i$ and $\bar{a}_i$ are as given in Table~\ref{aitable}.

\begin{table}[hbt]
\caption{The values of the coefficients $a_i$ and $\bar{a}_i$
for normal and inverted hierarchies
\label{aitable}}
\begin{center}
\footnotesize
\begin{tabular}{ccc}
\hline
& \multicolumn{2}{c}{Hierarchy} \\
& normal & inverted \\
\hline
$a_1$ & $P_H P_L$ &$ P_L$ \\
$a_2$ & $ P_H(1-P_L)$ &$ 1 - P_L$ \\
$a_3$ &$ 1 - P_H$ & 0 \\
$\bar{a}_1$ &$ 1 - \bar{P}_L$ & $\bar{P}_H ( 1 - \bar{P}  _L)$ \\
$\bar{a}_2$ &$\bar{P} _L$ &$\bar{P} _H \bar{P} _L$ \\
$\bar{a}_3$ & 0 & $1 -\bar{P} _H$ \\
\hline
\end{tabular}
\end{center}
\end{table}

The mass eigenstates arriving at the surface of the earth
oscillate in the earth matter.
Let $P_{ie}$ be the probability
that a mass eigenstate $\nu_i$ entering the
earth reaches the detector as a $\nu_e$.
The flux of $\nu_e$ at the detector is
\beq
F_e^D  = \sum_i P_{ie} F_i =  F_e^0 \sum a_i P_{ie} +
        F_x^0 (1 - \sum_i a_i P_{ie}) ~~,
\eeq
where we have
used the unitarity condition $\sum_i P_{ie} = 1$.
Thus, the $\nu_e$ flux at the detector can  be written as
\beq
F_e^D = p^D F_e^0 + (1 - p^D)  F_x^0 \quad {\rm with} \quad
p^D = \sum_i a_i P_{ie}~~.
\label{fed}
\eeq
Similarly, for the antineutrinos,  we can write
\beq
F_{\bar{e}}^D = \bar{p}^D F_{\bar{e}}^0 + 
(1 - \bar{p}^D)  F_x^0 \quad {\rm with} \quad
\bar{p}^D = \sum_i \bar{a}_i \bar{P}_{ie}~~.
\label{febard}
\eeq
The difference in the $\nu_e$ fluxes at the detector
due to the propagation in earth equals
\beq
F_e^D - F_e =   (p^D - p) (F_e^0 - F_x^0)~~,~~
F_{\bar{e}}^D - F_{\bar{e}} =   
(\bar{p}^D - \bar{p}) (F_{\bar{e}}^0 - F_x^0)~~.
\label{fed-fe}
\eeq

\subsection{Neutrinos}

In the case of normal hierarchy, using Table~\ref{aitable} we get
\beq
p^D - p =  P_H (P_{2e} - |U_{e2}|^2) ( 1 - 2 P_{L})
+ (P_{3e} - |U_{e3}|^2)(1 - P_H - P_H P_L)~~.
\label{pp-norm-ex}
\eeq
The second term in (\ref{pp-norm-ex}) can  be neglected.
Indeed, inside the earth, $\nu_3$ oscillates with a very small
depth:
\beq
P_{3e} - |U_{e3}|^2 \lsim
\left(\frac{2 E V_{earth}}{\Delta m^2_{atm}}\right)
\sin^2 2 \theta_{e3}~~,
\label{p3e}
\eeq
where $V_{earth}$ is the effective potential of
$\nu_e$ in the earth.
For neutrino energies of 5 -- 50 MeV,
we have
$2 E V_{earth}/ \Delta m^2_{atm} \lsim 10^{-2}$.
Moreover,
$\sin^2 2\theta_{e3} \leq 0.1$,
so that
$P_{3e} - |U_{e3}|^2 \leq 10^{-3}$.
Then (\ref{pp-norm-ex}) becomes
\beq
p^D - p \approx  P_H (P_{2e} - |U_{e2}|^2) ( 1 - 2 P_{L})~~.
\label{pp-norm}
\eeq
In general, when the signals from two detectors D1 and D2 are compared,
we get the difference of fluxes
\beq
F_e^{D1} - F_e^{D2} \approx P_H \cdot (1 - 2 P_L)
\cdot (P_{2e}^{(1)} - P_{2e}^{(2)})
\cdot (F_e^0 - F_x^0)~~,
\label{ff-norm}
\eeq
where $P_{2e}^{(1)}$ and  $P_{2e}^{(2)}$ are the
$\nu_e \leftrightarrow \nu_2$ oscillation probabilities  for the
detectors D1 and D2 correspondingly.

For inverted hierarchy, using Table~\ref{aitable} we get
\beq
p^D - p \approx (P_{2e} - |U_{e2}|^2) ( 1 - 2 P_{L})~~,
\label{pp-inv}
\eeq
so that
\beq
F_e^{D1} - F_e^{D2} \approx (1 - 2 P_L)
\cdot (P_{2e}^{(1)} - P_{2e}^{(2)})
\cdot (F_e^0 - F_x^0)~~.
\label{ff-inv}
\eeq

\subsection{Antineutrinos}

In the case of normal hierarchy, Table~\ref{aitable} leads to
\beq
\bar{p}^D - \bar{p} = (\bar{P}_{1e} - |U_{e1}|^2) (1 - \bar{P}_L) +
(\bar{P}_{2e}- |U_{e2}|^2) \bar{P}_L~~.
\label{ppbar-norm}
\eeq
Then 
we obtain the difference in the
fluxes at two detectors $D1$ and $D2$ as
\beq
F_{\bar{e}}^{D1} - F_{\bar{e}}^{D2} \approx (1 - 2 \bar{P}_L)
\cdot (\bar{P}_{1e}^{(1)} - \bar{P}_{1e}^{(2)}) \cdot 
(F_{\bar{e}}^0 - F_x^0)~~,
\label{ffbar-norm}
\eeq
where we have neglected the oscillations of $\bar{\nu}_3$
inside the earth.

For the inverted hierarchy,
\beq
\bar{p}^D - \bar{p} =
(\bar{P}_{1e}- |U_{e1}|^2) (1 - 2 \bar{P}_L) \bar{P}_H +
(\bar{P}_{3e} -|U_{e3}|^2) (1 - \bar{P}_H - \bar{P}_H \bar{P}_L)~~.
\label{ppbar-inv-ex}
\eeq
Since $\bar{\nu}_3$ oscillates inside the earth with a very
small depth (inequality (\ref{p3e}) is valid with $P_{3e}$
replaced by $\bar{P}_{3e}$),
the second term in (\ref{ppbar-inv-ex}) can be neglected to get
\beq
\bar{p}^D - \bar{p} = 
(\bar{P}_{1e}- |U_{e1}|^2) (1 - 2 \bar{P}_L) \bar{P}_H~~.
\label{ppbar-inv}
\eeq
Therefore, finally we get
\beq
F_{\bar{e}}^{D1} - F_{\bar{e}}^{D2} \approx \bar{P}_H \cdot
(1 - 2 \bar{P}_L) \cdot
(\bar{P}_{1e}^{(1)} - \bar{P}_{1e}^{(2)}) \cdot 
(F_{\bar{e}}^0 - F_x^0)~~.
\label{ffbar-inv}
\eeq

\section{``Factorized'' matter effects}
\label{sec:fact}

Note that the equations (\ref{ff-norm}), (\ref{ff-inv}), 
(\ref{ffbar-norm}) and (\ref{ffbar-inv}) have significant 
similarities. All of these
may be written in the factorized form
\beq
F^{D1} - F^{D2} = f_{flux} \cdot f_{star} \cdot f_{osc}~.
\label{fact}
\eeq
In this section, we shall discuss these factors and the insights they offer
on the extent of the observable earth matter effects.

\subsection{$f_{flux}$: the difference in initial fluxes}

The relevant difference in initial fluxes is
$$
f_{flux} = \left \{ \begin{array}{ll} 
(F_e^0 - F_x^0) &  \quad {\rm for} \quad \nu_e \\
(F_{\bar{e}}^0 - F_x^0) &  \quad {\rm for} \quad \bar{\nu}_e \\ 
\end{array} ~~ \right.  \quad .
$$
Since the $\nu_e$ and $\bar{\nu}_e$ spectra are softer than the $\nu_x$
spectrum, and the luminosities of all the spectra are
similar in magnitude \cite{janka-lum},
the term $f_{flux}$
is positive at low energies and becomes negative
at higher energies where the $\nu_x$ flux overwhelms the
$\nu_e$ ($\bar{\nu}_e$) flux. 
Therefore, the earth effect has a different sign
for low and high energies.
The observed effect can be significant in the energy range
where $\sigma \times f_{flux}$ is maximum (here $\sigma$ is
the cross section of the neutrino interactions).

\subsection{$f_{star}$: the factor
characterizing conversions inside the star}

The value of this factor depends only on the neutrino conversions 
occuring inside the star, and hence it also characterizes the
different scenarios:
\beq
f_{star} = \left \{ \begin{array}{ll} 
P_H (1 - 2 P_L)  & \quad {\rm normal~ hierarchy} \quad \nu_e \\
(1 - 2 P_L)  & \quad {\rm inverted~ hierarchy} \quad \nu_e \\
(1 - 2 \bar{P}_L) & \quad {\rm normal~ hierarchy} \quad \bar{\nu}_e \\
\bar{P}_H (1 - 2 \bar{P}_L) & 
\quad {\rm inverted~ hierarchy} \quad \bar{\nu}_e \\ 
\end{array}
\right.  \quad .
\eeq

Consider the $\nu_e$ conversions in the scenario of normal 
hierarchy.
In the limit of $U_{e3} \to 0$ and $P_H \to 1$,
(\ref{pp-norm}) reduces to the expression
for the earth effects in the case of two neutrino
mixing :
\beq
[p^D - p]_{2 \nu}= (P_{2e} -  |U_{e2}|^2) ( 1 - 2 P_{L}) ~~,
\label{d-n}
\eeq
which is equivalent to the one used in literature
in the context of day-night effect
for solar neutrinos. Therefore  (\ref{pp-norm}) can be
looked upon as the earth matter effect due to the
two neutrino mixing (\ref{d-n}) suppressed by a factor of $P_H$.
The mixing of the
third neutrino thus plays a major role, making the
expected earth effects in the case of supernova neutrinos
smaller than those expected in the case of solar
neutrinos for the same mixing scheme and in the same energy range.
If the $H$-resonance is completely adiabatic,
the earth effect vanishes:
all the $\nu_e$ produced
are converted to $\nu_3$ in the star, and the earth
matter effect on  $\nu_3$ is negligibly small (as we have
established through (\ref{p3e})).

Notice that the  expressions
for the earth matter effects in the case of the inverted mass hierarchy
(\ref{ff-inv}, \ref{ffbar-inv}) have the same form as the
expressions in the case of the normal mass
hierarchy  (\ref{ff-norm}, \ref{ffbar-norm}),
except for the factors of $P_H$ and $\bar{P}_H$. 
Indeed, since $P_H, \bar{P}_H \leq 1$, these act as suppression
factors for the earth matter effects on $\nu_e$ and $\bar{\nu}_e$,
in the case of normal and inverted hierarchy respectively.
Then the observation of significant earth matter effects on $\nu_e$ 
for normal hierarchy would imply $P_H \approx 1$, i.e. 
$\ue3sq \lsim 10^{-3}$. Similarly, 
significant earth matter effects on $\bar{\nu}_e$ 
for inverted hierarchy would imply $\bar{P}_H \approx 1$, i.e. 
$\ue3sq \lsim 10^{-3}$.

\subsection{$f_{osc}$: the difference in earth oscillation
probabilities}

This factor represents the difference in the neutrino 
conversions inside the earth due to the different distances
travelled by the neutrinos inside the earth 
before reaching the detectors.
$$
f_{osc} = \left \{ \begin{array}{ll}
P_{2e}^{(1)} - P_{2e}^{(2)} & \quad \quad {\rm for} \quad \nu_e \\
\bar{P}_{2e}^{(1)} - \bar{P}_{2e}^{(2)} & 
\quad \quad {\rm for} \quad \bar{\nu}_e \\
\end{array}
\right.  \quad .
$$

If the neutrino trajectory  crosses only the mantle of the earth,
one can use a constant density approximation which gives
\beq
P_{2e}^{(1)} - P_{2e}^{(2)}  \approx \sin 2\theta_{e2}^m
~ \sin (2\theta_{e2}^m - 2\theta_{e2})
~\left[ \sin^2 \left( \frac{\pi d_1}{l_m} \right) -
\sin^2 \left( \frac{\pi d_2}{l_m} \right) \right]~~.
\label{earth-nu}
\eeq
Here $\theta_{e2}^m$ and $l_m$ are  the mixing angle and
the oscillation length in the earth matter respectively, and
$d_i$ is the distance travelled by the
neutrinos inside the earth before reaching the detector $D_i$.
The first two terms on the right hand side of
(\ref{earth-nu}) are positive for the scenarios with the SMA and LMA
solutions, so that the
sign of $f_{osc}$ is the same as
the sign of the term inside the square bracket in
(\ref{earth-nu}).
For the scenario with the VO solution,
the earth matter effects are negligible:
for $\Delta m^2 \sim 10^{-10}$ eV$^2$, the mixing
in the earth matter is highly suppressed, since
$\sin 2\theta_{e2}^m$  is very small.

In the case of antineutrinos, the constant density
approximation gives
\beq
\bar{P}_{1e}^{(1)} - \bar{P}_{1e}^{(2)} \approx
- \sin 2\bar{\theta}_{e2}^m
~\sin (2\bar{\theta}_{e2}^m - 2\theta_{e2})
~\left[ \sin^2 \left( \frac{\pi d_1}{l_m} \right) -
\sin^2 \left( \frac{\pi d_2}{l_m} \right) \right]~~,
\label{earth-nubar}
\eeq
where $\bar{\theta}_{e2}^m$ is the mixing angle inside the earth
for the antineutrinos. For the antineutrino channel
$\bar{\theta}_{e2}^m < \theta_{e2} \ll 1$ for SMA solution and
$\bar{\theta}_{e2}^m$ is strongly suppressed by matter
in the VO case.
Therefore
the earth matter effects on the
$\bar{\nu}_e$ spectrum can be significant only for the
scenario with the LMA (as well as  LOW) solution. In this scenario,
$\sin 2\bar{\theta}_{e2}^m >0$ and
$\sin (2\bar{\theta}_{e2}^m - 2\theta_{e2}) <0$,
so that the sign of $f_{osc}$ is the
same as the sign of the oscillation term inside the square bracket
in (\ref{earth-nubar}).

If  neutrinos cross  both the mantle and the core, the
parametric enhancement of oscillations may occur, which leads to
the appearance of parametric peaks apart from the
peaks due to the MSW resonances in the core and the mantle
\cite{adls}. Correspondingly the factor $f_{osc}$ 
will be a more complicated function of the neutrino energy.

To summarize,
the earth matter effects on the $\nu_e$ spectrum can
be significant only for the scenarios with the SMA or LMA
or LOW solutions. Moreover, if the hierarchy is normal,
$\ue3sq \lsim 10^{-3}$ is needed in addition.
The effect on the $\bar{\nu}_e$ spectrum can only be significant for
the LMA scenario. Let us consider these cases in detail below.

In the case of SMA, the factor of $f_{osc}$ can be as large as
0.25 in the energy range of 20 -- 40 MeV, where
the term $f_{flux}$ is also significant. However, even in this
scenario, and with optimistic values of
$P_H, P_L$ and $\theta_{e2}$, the net effect
is only $\lsim 10$\%, which would be difficult to
disentangle from the uncertainties in the original fluxes
(see Fig.~7 in \cite{ds}).

\begin{figure}[htb]
\epsfxsize=30pc
\epsfbox{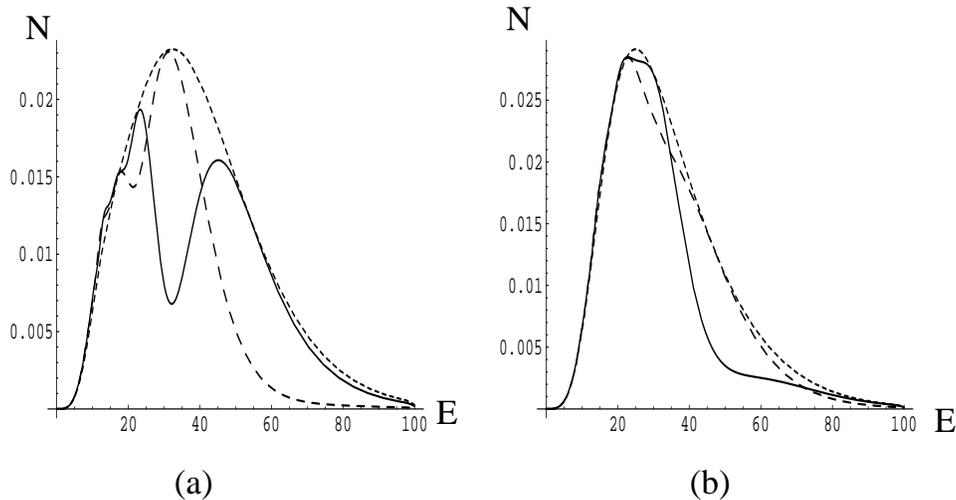}
\caption{The earth matter effects on (a) the $\nu_e$ spectrum
and (b) the $\bar{\nu}_e$ spectrum
for $P_H=1$ in the scheme with the LMA solution
($\Delta m^2 =2 \cdot 10^{-5}$ eV$^2,~\sin^2 2\theta_\odot= 0.9$)
and normal hierarchy.
The dotted, dashed and solid lines
show the spectra of the number of $\nu-N$ charged current events
when the distance travelled by the neutrinos
through the earth is
$d=0$ km, $d=4000$ km, and $d=6000$ km respectively.
\label{earth-lma}}
\end{figure}

In the case of LMA, the factor $f_{osc}$ can be as large as 0.3
in the energy range of 20 -- 50 MeV both for $\nu_e$ and
$\bar{\nu}_e$. 
Also, the transitions in the $L$ resonance layer are completely
adiabatic \cite{ds} for both neutrinos and antineutrinos,
so that the factor $(1-2P_L)$ as well as
$(1-2 \bar{P}_L)$ is equal to 1.
Indeed, the earth matter effects can be
large in both channels.
In Fig.~\ref{earth-lma}, we show the $\nu_e$ and $\bar{\nu}_e$
spectra for different distances travelled by the neutrinos
through the earth.

\begin{figure}[htb]
\epsfxsize=20pc
\epsfbox{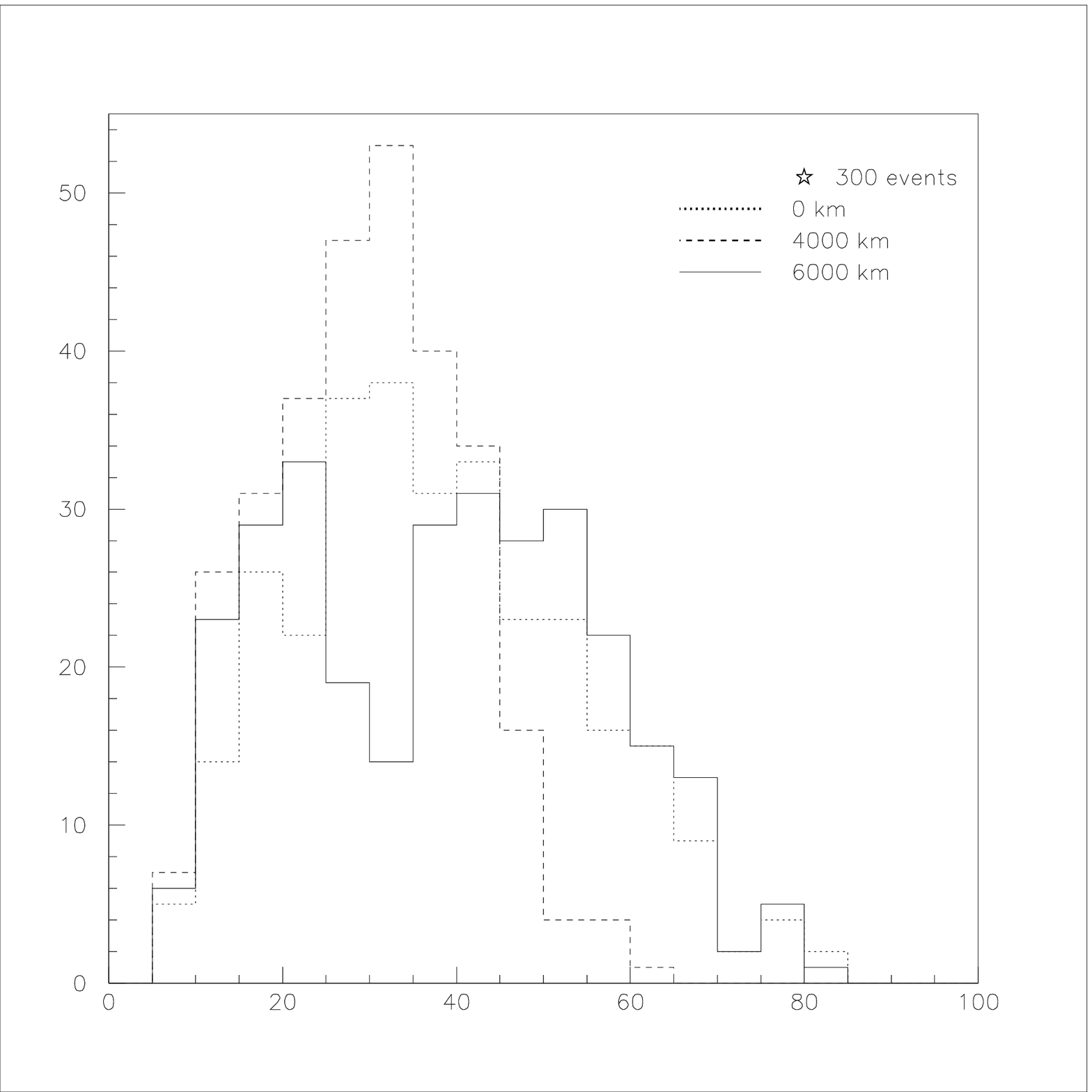}
\caption{The number of events (y-axis) of $\nu_e$ as a function of
energy in MeV (x-axis) for the LMA solution (parameters as in 
Fig~\ref{earth-lma}). The three spectra correspond to
$d=0,4000,6000$ km. The spectra are normalized so that the number
of events for each spectrum is 300.
\label{lma-monte}}
\end{figure}

Note that in the case of LMA, in some region of parameter space 
the final $\nu_e$ spectrum shows a spectacular dip \cite{ds,sato}, 
which cannot
be mimicked by any uncertainty in the original energy spectrum.
Thus, the earth matter effects can be demonstrated in a clean way,
independent of any model of supernova dynamics and by the observation
of signals in only one detector.
We demonstrate this case further through a Monte carlo simulation
of the final $\nu_e$ spectrum in LMA scheme with 300 events
in Fig.~\ref{lma-monte}. The dip is clearly observable in the 
spectrum with $d=6000$ km.

\section{Conclusions}

We have indicated how the
observation of earth matter effects on the $\nu_e$ and
$\bar{\nu}_e$ energy spectra from a supernova can 
help resolve ambiguties in the neutrino mass spectrum in a
model independent manner.

The observation of any earth matter effects rules out
the scenario with the VO solution.
If the earth matter effects are observed in the
neutrino channels but not in the antineutrino
channels, we either have the inverted mass hierarchy, or
the normal mass hierarchy with
$\ue3sq \lsim 10^{-3}$.

The observation of earth matter effects in the antineutrino
channel identifies the LMA solution.
In addition, if the effects are significant
in the neutrino channel also,
$\ue3sq \lsim 10^{-3}$ may be established.

The qualitative features of the earth matter effects
may be able to explain certain peculiar features of
the SN1987A supernova neutrino energy spectra
\cite{lunardini}, but the number of events is too
small to come to any definite conclusion. However,
a galactic supernova may provide us with a sufficient number of events
\cite{gandhi} to enable us to reconstruct the energy spectra
of $\nu_e$ and $\bar{\nu}_e$.

I would like to thank the organising committee of CICHEP'2001
for their hospitality during the Workshop.  
I would also like to thank A. Yu. Smirnov for discussions and insights 
during the collaboration for \cite{ds}.

This talk points out the salient features of the earth matter
effects on the supernova neutrino spectra.
For a more recent detailed study of the spectra expected at
various detectors, the reader is 
referred to \cite{latest}.

\end{document}